\documentclass[prb,aps,preprint]{revtex4}
\usepackage{graphicx}

\begin{document}

\title{Spin precession and inverted Hanle effect in a semiconductor near a finite-roughness ferromagnetic interface}
\author{S.P. Dash$^{1}$, S. Sharma$^{2,3}$, J.C. Le Breton$^{2,4}$, J. Peiro$^4$, H. Jaffr\`{e}s$^4$, J.-M. George$^4$, A. Lema\^{i}tre$^5$ and R. Jansen$^{3}$} \affiliation{$^1$\,Department of
Microtechnology and Nanoscience, Chalmers University of Technology, SE-41296,
G\"{o}teborg, Sweden.\\
$^2$\,Netherlands Foundation for Fundamental Research
on Matter (FOM), 3502 GA Utrecht, The Netherlands.\\
$^3$\,National Institute of Advanced Industrial Science and
Technology (AIST), Spintronics Research Center, Tsukuba, Ibaraki
305-8568, Japan.\\
$^4$\,Unit\'{e} Mixte de Physique CNRS-Thales, 91767 Palaiseau and Universit\'{e} Paris-Sud, 91405 Orsay, France,\\
$^5$\,Laboratoire de Photonique et Nanostructures, CNRS, 91460
Marcoussis, France.}



\begin{abstract}
Although the creation of spin polarization in various non-magnetic media via electrical spin
injection from a ferromagnetic tunnel contact has been demonstrated, much of the basic behavior is
heavily debated. It is reported here for semiconductor/Al$_2$O$_3$/ferromagnet tunnel structures
based on Si or GaAs that local magnetostatic fields arising from interface roughness dramatically
alter and even dominate the accumulation and dynamics of spins in the semiconductor. Spin
precession in the inhomogeneous magnetic fields is shown to reduce the spin accumulation up to
tenfold, and causes it to be inhomogeneous and non-collinear with the injector magnetization. The
inverted Hanle effect serves as experimental signature. This interaction needs to be taken into
account in the analysis of experimental data, particularly in extracting the spin lifetime $\tau_s$
and its variation with different parameters (temperature, doping concentration). It produces a
broadening of the standard Hanle curve and thereby an apparent reduction of $\tau_s$. For heavily
doped n-type Si at room temperature it is shown that $\tau_s$ is larger than previously determined,
and a new lower bound of 0.29 ns is obtained. The results are expected to be general and occur for
spins near a magnetic interface not only in semiconductors but also in metals, organic and
carbon-based materials including graphene, and in various spintronic device structures.

\end{abstract}

\maketitle



\section{INTRODUCTION}
\noindent The controlled creation of a non-equilibrium spin polarization in non-magnetic materials
is a central aspect of spintronics and plays a role in virtually all spin-based electronic
nanostructures \cite{wolf,chappert,fabianacta}. In the spin valve, the most well-known example of a
metallic spintronic device consisting of two ferromagnetic layers separated by a thin non-magnetic
metal, spin information can be transmitted between the two ferromagnets via the spin accumulation
in the spacer. This gives rise to giant magnetoresistance, exchange coupling, and allows one
ferromagnet to exert a torque on the other \cite{fertGMR,grunberg,slonczewski,berger}. In a spin
transistor, an example of a spin-based semiconductor device, spin information between ferromagnetic
source and drain is transmitted via a semiconductor channel \cite{dattadas,fertIEEE}, making it
possible to manipulate the spins during transit by a gate electric field. Understanding the physics
of spins in non-magnetic materials is thus crucial as it controls the overall behavior and
performance of spin-based nanostructures. Although spin polarization has been electrically created
in a variety of non-magnetic materials, mostly via spin-polarized tunneling from a ferromagnetic
contact
\cite{jedema1,jedema2,lou,jonker,erve,weiss,hueso,tombros,tran,hamaya,sasaki,dash,jansen2deg}, much
of the basic physics is not understood. The magnitude and sign of the induced polarization are
heavily debated\cite{tran,fertPRB,osipov,sham,chantis,dery}, the variation with bias voltage and
temperature is often puzzling \cite{lou,tran,dash,shiraishi,salis}, and unexpectedly short spin
lifetimes are observed, for instance in the range of a few hundred ps in graphene and doped Si at
room temperature \cite{tombros,dash}.
\subsection{Spins in proximity to a ferromagnetic interface}
\indent Because spintronic nanostructures combine different materials (ferromagnets with
non-magnetic metals, semiconductors, organic and carbon-based materials), a key question is to what
extent the proximity to interfaces influences the spin accumulation and the spin dynamics. Dipolar
fields from magnetic domain walls in a demagnetized Ni film have been reported to reduce the
spin-dephasing time of optically-excited carriers in GaAs \cite{korenev1,korenev2}, but the
associated increase of the Hanle line width ($\sim$ 1 Oe) is small. Spin precession is also known
to be affected by nuclear hyperfine fields \cite{merkulov,dzhioevprl,crowellnuclear}. These are not
related to the ferromagnetic interface and are typically relevant only at low temperature. In
contrast, we demonstrate here a much more general mechanism (present even at room temperature and
for homogeneously magnetized ferromagnetic electrodes) that has a surprisingly dramatic effect on
spin accumulation and spin dynamics of carriers in a non-magnetic medium near a magnetic interface.
Specifically, inhomogeneous magnetostatic fields arising from finite interface roughness are shown
to alter precession of spins in a semiconductor near the magnetic interface, dominate spin dynamics
up to surprisingly large external fields as large as 1 kOe, and reduce the spin accumulation up to
tenfold. We focus here on spin polarization created in semiconductors by injection of spins from a
ferromagnetic tunnel contact. However, the phenomena described here should occur irrespective of
the type of non-magnetic material or the method used to create the spin accumulation, although the
extent of the effect depends on the
details of the system.\\
\indent The magnetostatic fields near a ferromagnetic interface with finite roughness are sketched
in figure 1 for the case of a sinusoidal interface profile with period $\lambda$. The magnetization
of the ferromagnet is taken to lie in-plane and point strictly along the global interface
everywhere. This is a valid approximation for the soft magnetic thin films without significant
interface anisotropy that we use here, as their magnetization can easily be saturated in a small
in-plane magnetic field. While for an extended and perfectly flat, in-plane magnetized film the
magnetostatic field would be zero outside the ferromagnet, in the presence of finite roughness
there are local magnetostatic fields that penetrate into the non-magnetic medium and influence the
spins. Note that this is not only determined by the ferromagnet/tunnel barrier interface, but for
thin films also by the roughness of the top surface of the ferromagnet, due to the long range
nature of magnetic fields. The magnetostatic fields are inhomogeneous in magnitude and direction,
and change sign periodically. The magnitude of the fields scales with the roughness amplitude, and
is linearly proportional to the magnetization $M_s$ of the ferromagnet. The strength of the field
decays with distance $z$ from the interface on a length scale that, for periodic roughness, is set
\cite{demokritov} by the lateral roughness period $\lambda$. Under electrical spin injection from
the ferromagnetic contact, a spin accumulation $\Delta\mu=\mu^{\uparrow}-\mu^{\downarrow}$ is
induced, with $\mu^{\uparrow}$ ($\mu^{\downarrow}$) the electrochemical potential for electrons
with majority (minority) spin. In the absence of roughness, $\Delta\mu$ decays exponentially as a
function of distance $z$ from the injection interface (Fig. 1b), with a spin-diffusion length
$L_{SD}$. However, for finite roughness spin precession is altered significantly in the region
between z=0 and z=$\lambda$ where appreciable local magnetostatic fields exist, strongly reducing
$\Delta\mu$. Even if $\lambda$ is shorter than $L_{SD}$, interfacial depolarization reduces
$\Delta\mu$ over the full depth range (Fig. 1b) because spin diffusion connects all spins and
dictates that spatial variations in spin density cannot exist on a length scale much smaller than
$L_{SD}$. Hence, interfacial magnetostatic fields affect the spins to an effective depth of
$L_{SD}$. Also note that by spin-polarized tunneling into the ferromagnet one probes the value of
$\Delta\mu$ at $z=0$, where the reduction is strongest as the spin accumulation right at the
interface is most directly affected by the local magnetostatic fields.
\section{TUNNEL CONTACTS AND MEASUREMENTS}
\indent We describe results for tunnel contacts on two different semiconductors (Si and GaAs). The
device fabrication and electrical measurement techniques have been described previously
\cite{tran,dash}. In brief, tunnel contacts of Si/Al$_2$O$_3$/FM have been prepared \cite{dash} by
evaporation in ultrahigh vacuum using different ferromagnets (FM) on n-type as well as p-type Si
substrates (carrier density and resistivity of 1.8$\times$10$^{19}$ cm$^{-3}$ and 3 m$\Omega$cm at
room temperature for n-type Si with As doping, and 4.8$\times$10$^{18}$ cm$^{-3}$ and 11
m$\Omega$cm at room temperature for p-type Si with B doping). The GaAs/Al$_2$O$_3$/Co structures
\cite{tran} are grown by sputtering on n-type GaAs epilayers with a doping concentration of
5$\times$10$^{18}$ cm$^{-3}$ with a 15 nm heavily doped surface region (2$\times$10$^{19}$
cm$^{-3}$). All measurements are performed on contacts having dimensions of 100$\times$200
$\mu$m$^2$ (Si) and 15$\times$196 $\mu$m$^2$ (GaAs) in the so-called three-terminal geometry
\cite{tran,dash}, probing the spin accumulation near a single ferromagnetic tunnel interface, thus
using the same contact for spin injection and detection. Roughness characterization is presented in
appendix D.
\section{RESULTS}
\subsection{Spin precession in silicon near a ferromagnetic interface}
\indent When spin-polarized electrons tunnel from the ferromagnet into the semiconductor, the
injected spins initially point along the magnetization direction of the ferromagnet, taken to be
along $x$. Ideally, in the absence of an external applied magnetic field $B^{ext}$ there is no
Larmor spin precession, and a static, non-equilibrium spin accumulation is induced. The local
magnetostatic fields $B^{ms}(x,y,z)$ modify this simple picture. Spins are precessing in the total
magnetic field that is composed of $B^{ext}$ and $B^{ms}(x,y,z)$. Since the latter is spatially
inhomogeneous in direction and amplitude, the axis of spin precession and the precession frequency
become spatially inhomogeneous. A full account of the consequences is given in
the model section, after having described the experimental data.\\
\indent The spin accumulation is probed by establishing a constant tunnel current I across the
semiconductor/Al$_2$O$_3$/FM tunnel contact, and measuring the change in voltage $\Delta$V across
that same tunnel contact as a function of $B^{ext}$. Since \cite{fertIEEE,osipov,dery,brataas} the
tunnel resistance is directly proportional to $\Delta\mu$ (i.e., $\Delta$V=TSP$\times \Delta\mu$/2
with TSP the tunnel spin polarization associated with the Al$_2$O$_3$/FM interface) and $\Delta\mu$
is reduced by spin precession, the value of $\Delta$V and its variation with $B^{ext}$ provide
information about the spin dynamics. We start with n-type Si and conventional Hanle measurements
(figure 2, left panel), with $B^{ext}$ applied along the z-axis (perpendicular to the interface and
to the injected spins). A typical Hanle curve is observed, with a maximum voltage (and hence
$\Delta\mu$) at $B^{ext}$=0, and a gradual reduction with increasing external field due to spin
precession. This is similar to Hanle data obtained previously \cite{dash}, establishing that a
non-equilibrium spin accumulation in the Si is induced by the injection of the spin-polarized
tunnel current. Control experiments have previously excluded artifacts not related to spin
injection \cite{dash}. Previous work has also unambiguously established that the room-temperature
spin polarization exists {\em in the bulk bands} of the Si rather than being enhanced by localized
interface states (see the specific experiments reported in Fig. 3 of Ref. \onlinecite{dash}, and
the observation of circularly-polarized electroluminescence originating from 300 nm away from the
injection interface in Si-based spin light emitting diodes \cite{spinled}). Despite this, we
observe, similar to previous work \cite{dash}, spin signals for different ferromagnets (see below)
in the range of 1-10 k$\Omega\mu$m$^2$ and thus larger than expected from theory, as noted
before\cite{dash}. The origin of this disagreement is still under discussion, but since enhancement
by localized states has already been ruled out \cite{dash}, there must be other enhancement factors
that are not yet incorporated in existing theory. This is beyond the scope of the present work,
which is concerned with the generic phenomena that affect the spin precession near a ferromagnetic
interface and thereby the shape of the Hanle curve. Therefore we will here not discuss the factors
that determine the overall magnitude of $\Delta\mu$, and show only normalized data. We did not find
any correlation between the overall signal magnitude and the shape of the curves.\\
\indent Let us now focus on the features that are due to the proximity of the interface with the
ferromagnet. We find that the width of the Hanle curve depends on the ferromagnet used, i.e., the
width increases from Ni, to Ni$_{80}$Fe$_{20}$, to Co, to Fe, with a half-width-at-half-maximum
(HWHM) of 200, 400, 710, and 1030 Oe, respectively. Conventionally, the Hanle curves are described
\cite{fabianacta,dash} by a Lorentzian given by $\Delta\mu(B)$=$
\Delta\mu(0)/(1+(\omega_L\cdot\tau_s)^2)$, where $\tau_s$ is the spin lifetime and $\omega_L$ is
the Larmor frequency ($\omega_L$=$g{\mu}_{B}B$/$\hbar$, where $g$ is the Land\'e g-factor,
${\mu}_{B}$ the Bohr magneton and $\hbar$ Planck's constant divided by 2$\pi$). The width of the
Hanle curve is then set solely by parameters of the semiconductor ($\tau_s$ and $g$), inconsistent
with our data. We attribute the experimental trend to modification of the spin dynamics near the FM
interface due to local magnetostatic fields that arise for finite roughness. As shown in the model
section below, this produces an artificial broadening of the Hanle curve that depends on the
direction and magnitude of $B^{ms}$, which in turn is proportional to the magnetization ($M_s$) of
the FM. Indeed ${\mu}_{0}M_s$ at room temperature increases from 0.6 T for Ni, to 0.9 T for
Ni$_{80}$Fe$_{20}$, to 1.8 T for Co, and to 2.2 T for Fe.
\subsection{Inverted Hanle effect}
\indent The above interpretation is proved by the following phenomenon, hereafter referred to as
the {\em inverted} Hanle effect. It denotes the increase of the spin polarization in an applied
(longitudinal) magnetic field (in analogy with the term Hanle effect, which gives a reduction of
the spin polarization in an applied (transverse) magnetic field). If $B^{ext}$=0, the spin
accumulation will be reduced by precession in the $y$ and $z$ components of the local magnetostatic
fields, which are orthogonal to the injected spins for a ferromagnet with magnetization along $x$.
If now a non-zero $B_x^{ext}$ along $x$ is added and increased, the total magnetic field (vector
sum of $B^{ms}$ and $B_x^{ext}$) rotates into the direction of the magnetization, thus reducing the
angle between the injected spins and the axis of precession. The precession is suppressed, and an
increase in the spin accumulation is expected as a function of $B_x^{ext}$. Indeed, the data in
figure 2 shows exactly this inverted Hanle effect for all FM electrodes. The smallest voltage (and
hence $\Delta\mu$) is obtained for $B_x^{ext}$=0, while at large $B_x^{ext}$ the voltage across the
contact saturates as spin precession in the local magnetostatic fields is fully eliminated. The
saturation occurs at a larger field value for the ferromagnet with larger M$_s$, consistent with
the outlined scenario. No dependence on the direction of the field in the x-y plane was observed,
as expected for poly-crystalline magnetic films for which roughness-induced magnetostatic fields
should be isotropic. We conclude that application of an external in-plane magnetic field leads to a
recovery of the spin accumulation, reaching the ideal value (that would be obtained without any
precession) for large enough $B_x^{ext}$. The "true" value of the spin accumulation is thus given
by the difference between the saturation signal of the inverted Hanle curve (large $B_x^{ext}$) and
the minimum of the signal of the conventional Hanle curve with $B^{ext}$ along $z$. This difference
has been normalized to 1 for all data presented. Importantly, the precession in local magnetostatic
fields causes a significant reduction of the spin accumulation, with $\Delta\mu$ at $B^{ext}$=0
varying from 10\% to 31\% of the ideal value.\\
\indent Note that an inhomogeneous spin accumulation can in principle also arise if the interface
magnetization does not point along the global interface plane everywhere, as this would lead to
inhomogeneity in the orientation of the spins that are injected. However, the in-plane magnetic
coercivity of the magnetic films used here is 5 - 30 Oe and the films do not have any significant
interface anisotropy. Therefore, the ferromagnet is homogeneously and fully magnetized along the
external in-plane field well below 100 Oe. Hence, the spin injection is homogeneous and does not
change for fields between 100 Oe and several kOe where the signal variation due to the inverted
Hanle effect is observed. Even if some slight deviation of the interface magnetization from
strictly in-plane were present, this cannot account for the strongly reduced spin accumulation that
is observed. This would require injection of carriers with spin pointing almost along the interface
normal. This is not plausible, and inconsistent with the magnetic behavior of magnetic tunnel
junctions prepared from the same materials \cite{min}.

\indent Qualitatively similar results are obtained for tunnel contacts on p-type Si (figure 3). For
all ferromagnets, a Hanle signal is observed at room temperature, consistent with our previous work
on the creation of spin polarization in p-type Si \cite{dash}. For increasing M$_s$ the width of
the Hanle curve increases, with HWHM of 200 Oe (Ni), 210 Oe (Ni$_{80}$Fe$_{20}$), 515 Oe (Co) and
950 Oe (Fe), although the difference between Ni$_{80}$Fe$_{20}$ and pure Ni is small. For all
devices an inverted Hanle curve is observed too, with a width and saturation field that increases
systematically for FM electrodes with larger M$_s$. The induced $\Delta\mu$ at $B^{ext}$=0 is about
27\% of the ideal value, but with less variation compared to the data for
n-type.\\
\indent In principle one can still fit the Hanle curves with a Lorentzian and extract a time
constant (given as labels in the left panels of Fig. 2 and 3). However, it should be treated as an
effective time or a lower limit to the spin lifetime in the semiconductor, because interface
magnetostatic fields are present and cause artificial broadening of the Hanle curve. Experimentally
this situation is easily recognized if an inverted Hanle effect is observed. Nevertheless, the
lower bound for the spin lifetime in the n-type Si we obtain (285 ps, Ni electrode) is already an
improvement by a factor of two compared to previous work with Ni$_{80}$Fe$_{20}$ electrodes
\cite{dash}, and the true spin lifetime is expected to be larger.
\subsection{Spin precession in GaAs near a ferromagnetic interface}
\indent A similar set of experiments was carried out on GaAs/Al$_2$O$_3$/Co tunnel junctions at
T=10 K (Figure 4). A Hanle signal is observed for $B^{ext}$ along $z$, establishing that a
non-equilibrium spin accumulation is created, although it has previously been proposed \cite{tran}
that the spins in these structures may accumulate primarily in localized states at or near the
interface. Of course, spins in localized states also feel the magnetostatic fields from the nearby
FM, consistent with the observation of the inverted Hanle effect (pink curves). The HWHM of the
Hanle curve is 1070 Oe, slightly larger compared to Si contacts with Co electrodes. The difference
may be due to a different amplitude of the roughness, and/or the larger magnetization at low
temperature. The effective time constant extracted from a fit to a Lorentzian is 1/$\omega$=55 ps,
assuming a g-factor of 2 for electrons in localized interface states. The induced $\Delta\mu$ at
$B^{ext}$=0 is 12\% of the maximum spin accumulation. It should be noted that for spin accumulation
in localized states in GaAs/Al$_2$O$_3$/Co structures at low temperature, we cannot completely rule
out that the behavior is caused by local magnetic (hyperfine) fields from nuclear spins
\cite{merkulov}, as previously studied with optical techniques in Voigt and Faraday geometry
\cite{dzhioevprl}. However, given the results of the Si devices, it is highly likely that local
magnetostatic fields arising from roughness are at the very least
partly responsible for the behavior of the GaAs devices.\\
\indent Additional insight is obtained from data at larger magnetic field (Fig. 4, bottom panel).
When $B_z^{ext}$ is increased, the spin signal is first reduced due to the Hanle effect, but then
sharply increases when the magnetization of the FM rotates out of plane, followed by a saturation
of the spin accumulation at large fields when the magnetization, and hence the spins in the GaAs,
are fully aligned with $B_z^{ext}$. Precession is then absent and the maximum $\Delta\mu$ is
obtained. The value of $\Delta\mu$ thus achieved should be identical to the saturation value of the
inverted Hanle curve, for which magnetization, spins in the GaAs and $B^{ext}$ all point along the
$x$-axis and precession is absent too. A difference is however observed, attributed to anisotropy
of the tunneling process \cite{TAMR,TAMRfabian,TASP}. Apart from some quantitative differences, the
results for GaAs and Si based devices are remarkably similar.
\section{MODEL}
\indent First, we briefly address an important difference with so-called orange-peel coupling that
exists between {\em two} ferromagnets in layered structures with finite roughness
\cite{kools,parkin}. Due to the exchange interaction in a ferromagnet, it feels only an average
magnetostatic field from the other ferromagnet, reducing the effective coupling field to a few tens
of Oe. In contrast, in a non-magnetic semiconductor the spins in different locations near the
ferromagnetic electrode can precess independently, and sense the full local strength of the
magnetostatic field, rather than an average. Hence, the relevant magnetic field scale for spins
accumulating in a non-magnetic material near a ferromagnet with finite roughness is
much larger than that of orange-peel coupling.\\
\indent The model that captures the basic physics of spin accumulation and precession near a
ferromagnetic interface and correctly describes the salient experimental behavior starts from the
equation \cite{fabianacta,opticalorientation} for spin dynamics of an ensemble of spins in a
non-magnetic host:
\begin{equation}
\frac{\partial {\bf S}}{\partial t} = {\bf S} \times {{\bf
\omega}_L} + D\nabla^2{\bf S} - \frac{\bf S}{\tau_s}.
\end{equation}
where {\bf S} is the spin density and ${{\bf
\omega}_L}=(\omega_x,\omega_y,\omega_z)=(g\mu_B/\hbar)\,(B_x,B_y,B_z)$. Terms on the right-hand
side describe, respectively, spin precession, spin diffusion (D the diffusion constant), and spin
relaxation. Spin drift has been neglected. We seek a solution for a homogeneous $B^{ext}$ plus
inhomogeneous magnetostatic fields near the FM interface: $B_i=B_i^{ext}+B_i^{ms}(x,y,z)$, with i=x,y,z.\\
\indent In the limit where the spin-diffusion length $L_{SD}$ is small compared to the roughness
period $\lambda$, the spin-diffusion term in eqn. (1) can be neglected. This provides an analytical
solution that is strictly correct when electrons are sufficiently localized for gradients in the
spin density to be sustained on the length scale of $\lambda$. This applies to the case of spin
accumulation in localized states (as in the GaAs devices \cite{tran}). It is not strictly valid for
mobile electrons since spin diffusion tends to average out the inhomogeneity of the spin density
(in our Si devices $L_{SD}$ is \cite{dash} at least a few 100 nm, while $\lambda$ is estimated to
be 20-60 nm, see appendix D). The net result is a more homogeneous spin density, but with a reduced
value. Although a rigorous, but cumbersome, numerical treatment including spin diffusion can be
done, we can expect that the value of the spin accumulation with spin diffusion is comparable to
spatial average of the inhomogeneous spin density that is calculated without spin diffusion. We
therefore average the spin-density over the x-y plane, finding that the basic experimental trends
of the Si and GaAs devices are reproduced. Without spin diffusion, the general steady state
solution for the x, y and z components of the spin density is \cite{opticalorientation} (see also
appendix A):
\begin{eqnarray}
S_x = S_0 \left\{\frac{\omega_x^2}{\omega_L^2} +
\left(\frac{\omega_y^2 +
\omega_z^2}{\omega_L^2}\right)\left(\frac{1}{1 + (\omega_L
\tau_s)^2}\right)\right\}\\
S_y = S_0 \left\{\frac{\omega_x \omega_y}{\omega_L^2} -
\left(\frac{\omega_x\omega_y}{\omega_L^2}+\omega_z\tau_s\right)\left(\frac{1}{1
+ (\omega_L \tau_s)^2}\right)\right\}\\
S_z = S_0 \left\{\frac{\omega_x \omega_z}{\omega_L^2} -
\left(\frac{\omega_x\omega_z}{\omega_L^2}-\omega_y\tau_s\right)\left(\frac{1}{1
+ (\omega_L \tau_s)^2}\right)\right\}
\end{eqnarray}
where $\omega_L^2=\omega_x^2 + \omega_y^2 + \omega_z^2$ and
$\omega_i=\omega_i^{ext}+ \omega_i^{ms}(x,y,z)$. Importantly, as
$\omega^{ms}$ is spatially inhomogeneous, the spin density is too.
Secondly, while the injected tunnel electrons have spin along the
x axis, for non-zero $B^{ms}$ the steady state spin density has x,
y and z components and is thus generally non-collinear with the
magnetization of the ferromagnetic injector (pointing strictly
along x). Third, without external field there is no suppression of
the spin polarization if $\omega_y^{ms}=\omega_z^{ms}=0$
($S_x=S_0$ and $S_y=S_z=0$), whereas $S_x<S_0$ in the presence of
magnetostatic fields with components orthogonal to the injected
spins (i.e., when $\omega_y^{ms}\neq0$ and/or
$\omega_z^{ms}\neq0$). Hereafter we shall focus on the $S_x$
component, since in electrical detection using the same
ferromagnetic tunnel contact only this component is relevant (the
tunnel resistance is proportional to the projection of the spin
accumulation onto the detector magnetization).\\
\indent To evaluate $B^{ms}$ of a FM with finite roughness, we describe it as a 2-dimensional
square array of magnetic dipoles pointing along x, and calculate the magnetostatic fields (see
figure 5). This gives an inhomogeneous pattern with all three field components present.
Alternatively, for 1-dimensional roughness an exact expression \cite{nogaret} of $B^{ms}$ in terms
of roughness amplitude and $M_s$ is given in appendix C. From this, and the measured roughness of
our structures (appendix D), we find that the strength of the magnetostatic fields
can easily be in the range of 1 kOe to 100 Oe up to a distance of 10 nm from the interface.\\
\indent Figure 6 shows spatial maps of $S_x$ obtained from eqn.
(2) using the magnetostatic fields at a distance of 5 nm from a
square array of dipoles, and $\tau_s$=1 ns. For $B^{ext}$=0, the
left panel shows regions with strongly reduced spin density (blue)
due to precession in the local (y and z components of the)
magnetostatic fields, and regions where precession is absent and
the maximum spin accumulation is present (red). When an external
magnetic field is added along z (Hanle configuration, top row) the
precession is enhanced everywhere and the spin density is
gradually reduced. In contrast, when $B^{ext}$ is applied along x
(inverted Hanle configuration, bottom row), everywhere the spin
density increases towards its maximum value (red) as precession in
$B^{ms}$ is suppressed. By averaging these maps over the x-y
plane, we obtain the variation of the average spin density as a
function of $B^{ext}$ (right two panels). This qualitatively
reproduces the experimental data: (i) the spin density at
$B^{ext}$=0 is reduced from its maximum, (ii) there is an inverted
Hanle effect, (iii) the width of the conventional Hanle curve is
broadened as compared to the situation without magnetostatic
fields, which would produce a Lorentzian with $\tau_s$=1 ns (shown
in green, with amplitude adjusted for easy comparison) and, (iv)
for increasing amplitude of $B^{ms}$ (larger dipole moment, bottom
panel), the width of the Hanle curve increases, and the inverted
Hanle curve and the reduction of the spin density at $B^{ext}$=0
become more pronounced. We conclude that, despite the neglect of
spin diffusion, the model agrees well with the experimental
observations and captures the basic physics.\\
\indent Above we have included $B^{ms}$ only in $\omega_L$ of the precession term of eqn. (1),
without changing $\tau_s$ in the last term. That is, we have modelled the phenomenon as being due
to changes in the axis and frequency of the (locally) {\em coherent} precession of the ensemble
spin polarization, modifying the measured time average of the spin density, and leading to
artificial broadening of the Hanle curve and thereby an {\em apparent} shortening of the spin
lifetime. In addition, the spatial inhomogeneity of the magnetostatic fields leads to decoherence
and further broadening. Let us now consider whether the inhomogeneous magnetic fields have an
effect on $\tau_s$. For localized electrons there is no effect on $\tau_s$. However, mobile
electrons near a FM interface moving through a spatially inhomogeneous magnetostatic field
experience this as a field fluctuating in time. This is distinct from D'yakonov-Perel' spin
relaxation, where the fluctuation is due to changes of the momentum, rather than the location in
real space that is relevant here. The associated time scale is given by
$\tau^{ms}=\lambda/4{\upsilon}$, where $\upsilon$ is the carrier velocity and $\lambda/4$ the
length scale over which the field changes significantly. Since electrons with different
trajectories acquire a different spin precession phase and transport is random, this causes
irreversible dephasing of the ensemble spin. Considering an electron moving parallel to the
interface and typical parameters ($\lambda < 100$ nm and $\upsilon=10^{5}$ m/s for electrons in
Si), $\tau^{ms}$ is below 1 ps and thus smaller than the spin-precession period for practical
fields ($1/\omega_L\geq5$ ps for $B\leq1$ T). Hence, we are in the regime of motional narrowing
\cite{zutic} and the associated spin-dephasing time is given \cite{zutic} by
$T_2^{ms}=1/\Omega_{av}^2\,\tau^{ms}$, where $B_{av}=\hbar\Omega_{av}/g\mu_B$ is the average
amplitude of the magnetostatic field. We thus have $1/T_2 = 1/T_2^{bulk} + 1/T_2^{ms}$, where
$T_2^{bulk}$ is the regular spin-dephasing time in the absence of local magnetostatic fields. For
$B_{av}\leq100$ mT and $\tau^{ms}=1$ ps we obtain $T_2^{ms}\geq3$ ns. This is larger than the spin
lifetimes we observe, and we therefore described the spin dynamics with a single spin lifetime,
including the magnetostatic fields only in the coherent precession term of eqn. (1). In other
situations, especially when $T_2^{bulk}$ is large, this source of dephasing may be of importance or
even become limiting.
\section{IMPLICATIONS}
\indent Perhaps the most immediate implication relates to the spin lifetime in Si, which was
previously \cite{dash} extracted from Hanle data to be about 140 ps for heavily doped n-type Si at
room temperature. We observe a clear inverted Hanle effect, the experimental signature that the
conventional Hanle curve is artificially broadened by interfacial magnetostatic fields from the FM,
such that a fit to a Lorentzian will underestimate the spin lifetime. Indeed, a new lower bound to
the spin lifetime for the n-type Si at room temperature was determined here (285 ps), and the
actual spin lifetime must still be larger than that. The artificial broadening may also obscure the
intrinsic variation of the spin lifetime with parameters such as temperature and doping
concentration, and should thus be considered to allow a meaningful discussion of trends. Similar
implications may be expected for other material systems, particularly when spins accumulate close
to the FM, such as in a single layer of graphene. More generally, the phenomena described here
shall appear for spin accumulation near ferromagnetic interfaces created by any means (such as
optical injection, electrical injection by tunneling, diffusive or ballistic transport, or via spin
Hall and other spin-orbit effects), in different device geometries (2-, 3-, and non-local
4-terminal devices), and for various non-magnetic materials (metals, semiconductors, organic and
carbon-based systems). The roughness-induced local magnetostatic fields and the resulting
inhomogeneity of the spin accumulation and precession should be taken into account in the analysis
of spin transport and dynamics, and may affect the properties and performance of spintronic
devices.

\begin{appendix}
\section{Steady-state spin accumulation for arbitrary magnetostatic field}
\indent A local magnetostatic field $B^{ms}$ arising from interface roughness adds to the external
applied magnetic field $B^{ext}$ and thereby changes the local axis of coherent spin precession, as
well as the precession frequency. To describe this, we start from the equation \cite{fabianacta}
for spin dynamics of an ensemble of spins in a non-magnetic host:
\begin{equation}
\frac{\partial {\bf S}}{\partial t} = {\bf S} \times {{\bf \omega}_L} + D\nabla^2{\bf S} -
\frac{\bf S}{\tau_s}.
\end{equation}
where {\bf S} is the spin density and ${{\bf
\omega}_L}=(\omega_x,\omega_y,\omega_z)=(g\mu_B/\hbar)\,(B_x,B_y,B_z)$. Terms on the right-hand
side describe, respectively, spin precession, spin diffusion (D the diffusion constant), and spin
relaxation. We have neglected spin drift. The x, y and z components of the spin density are
explicitly written as:
\begin{eqnarray}
\frac{\partial S_x}{\partial t} = S_y \omega_z - S_z \omega_y +
D\nabla^2 S_x - \frac{S_x}{\tau_s}\\
\frac{\partial S_y}{\partial t} = S_z \omega_x - S_x
\omega_z + D\nabla^2 S_y - \frac{S_y}{\tau_s}\\
\frac{\partial S_z}{\partial t} = S_x \omega_y - S_y \omega_x + D\nabla^2 S_z - \frac{S_z}{\tau_s}
\end{eqnarray}
If spin diffusion can be neglected (for spin-diffusion length $L_{SD}$ much smaller than the period
$\lambda$ of the roughness), and the boundary conditions at $t=0$ are:
\begin{eqnarray}
S_x(t=0) = A\\
S_y(t=0) = 0\\
S_z(t=0) = 0
\end{eqnarray}
then the analytic solutions for arbitrary magnetic field are given by:
\begin{eqnarray}
S_x(t) = A \left\{\frac{\omega_x^2 + (\omega_y^2 + \omega_z^2)cos(\omega_L t)}{\omega_L^2}\right\}exp(-t/\tau_s)\\
S_y(t) = A \left\{\frac{\omega_x \omega_y - \omega_x\omega_y cos(\omega_L t) - \omega_L\omega_z sin(\omega_L t)}{\omega_L^2}\right\}exp(-t/\tau_s)\\
S_z(t) = A \left\{\frac{\omega_x \omega_z - \omega_x\omega_z cos(\omega_L t) + \omega_L\omega_y
sin(\omega_L t)}{\omega_L^2}\right\}exp(-t/\tau_s)
\end{eqnarray}
with $\omega_L^2=\omega_x^2 + \omega_y^2 + \omega_z^2$. These expressions describe the time
evolution of a packet of spins initially polarized along the x-axis at $t=0$. The steady state spin
polarization under continuous injection is proportional to the time integral $\int_0^\infty S_i(t)
dt$, which yields:
\begin{eqnarray}
S_x = S_0 \left\{\frac{\omega_x^2}{\omega_L^2} + \left(\frac{\omega_y^2 +
\omega_z^2}{\omega_L^2}\right)\left(\frac{1}{1 + (\omega_L
\tau_s)^2}\right)\right\}\\
S_y = S_0 \left\{\frac{\omega_x \omega_y}{\omega_L^2} -
\left(\frac{\omega_x\omega_y}{\omega_L^2}+\omega_z\tau_s\right)\left(\frac{1}{1
+ (\omega_L \tau_s)^2}\right)\right\}\\
S_z = S_0 \left\{\frac{\omega_x \omega_z}{\omega_L^2} -
\left(\frac{\omega_x\omega_z}{\omega_L^2}-\omega_y\tau_s\right)\left(\frac{1}{1 + (\omega_L
\tau_s)^2}\right)\right\}
\end{eqnarray}
where $S_0$ is the spin polarization in the absence of any magnetic field, and
$\omega_i=\omega_i^{ext}+ \omega_i^{ms}(x,y,z)$. Eqn. (11) can be written in terms of a solid angle
$\theta$ between injected spins and magnetic field vector, as done previously
\cite{opticalorientation} for optical excitation:
\begin{equation}
S_x = S_0 \left\{cos^2(\theta) + \left(\frac{sin^2(\theta)}{1 + (\omega_L \tau_s)^2}\right)\right\}
\end{equation}
Without an external applied magnetic field, the spin polarization is determined exclusively by the
local magnetostatic fields due to roughness:
\begin{equation}
S_x = S_0 \left\{\frac{(\omega_x^{ms})^2}{(\omega_L^{ms})^2} + \left(\frac{(\omega_y^{ms})^2 +
(\omega_z^{ms})^2}{(\omega_L^{ms})^2}\right)\left(\frac{1}{1 + (\omega_L^{ms}
\tau_s)^2}\right)\right\}
\end{equation}
where $(\omega_L^{ms})^2=(\omega_x^{ms})^2 + (\omega_y^{ms})^2 + (\omega_z^{ms})^2$. Note that the
reduction of the spin polarization depends, in general, on the strength as well as on the
orientation of the local magnetostatic fields. However, in the limit $\omega_L^{ms}\tau_s>>1$, only
the orientation of the field is relevant and $S_x$ becomes independent of the field strength (and
hence independent of the magnetization of the ferromagnet).

\section{Hanle curves in the presence of local magnetostatic fields}
\indent The effect of the different components of the local magnetostatic field on the Hanle curves
was calculated from eqn. (11) using, for illustrative purposes, magnetostatic fields B$^{ms}$
pointing purely along either the x, y or z-axis. The field strength is taken to have a periodic
spatial variation $\omega_i^{ms}=\omega_{0}\,cos(2\pi x/\lambda$), and the spin polarization was
averaged in space over a full period $\lambda$. The resulting Hanle curves (B$^{ext}$ along z) and
inverted Hanle
curves (B$^{ext}$ along x) are shown in figure \ref{figS1}. We see that for:\\
\\
(i) {\it B$^{ms}$ along x, parallel to the injected spins:}\\
The Hanle curve is broadened, but there is no inverted Hanle signal and no reduction of the spin
accumulation at zero external
field.\\
\\
(ii) {\it B$^{ms}$ along y, orthogonal to the injected spins:}\\
The Hanle curve is broadened, there is an inverted Hanle signal,
and a reduction of the spin accumulation at zero external field.\\
\\
(iii) {\it B$^{ms}$ along z, orthogonal to the injected spins:}\\
There is an inverted Hanle signal and a reduction of the spin accumulation at zero external field,
while the Hanle curve is broadened, as well as split into two components, corresponding to
locations with B$_z^{ms}$ aligned or anti-aligned with the external B$_z^{ext}$ field.

\section{Local magnetostatic fields near a ferromagnet with 1-dimensional roughness}
\indent The pattern and magnitude of the local magnetostatic fields for a ferromagnet with
1-dimensional roughness can be obtained via a Fourier transform \cite{nogaret}. Taking the surface
height to vary along the x-axis with period $\lambda$, a square height profile with peak-to-peak
height $h$, and magnetization pointing along the x-direction, we have\cite{nogaret}:
\begin{eqnarray}
B_x^{ms}(x,z) = \mu_0 M_s \left(\frac{h}{2}\right) \sum_{n=1}^{\infty} q_n \, F(q_n) \exp(-q_n z)\,sin(q_n x-\pi/2),\\
B_y^{ms}(x,z) = 0,\\
B_z^{ms}(x,z) = \mu_0 M_s \left(\frac{h}{2}\right) \sum_{n=1}^{\infty} q_n \, F(q_n) \exp(-q_n z)
\,cos(q_n x-\pi/2),
\end{eqnarray}
where $q_n=2\pi n/ \lambda$, and
\begin{equation}
F(q_n) = \frac{sin(q_n \lambda /4)}{(q_n \lambda /4)} \frac{sinh(q_n h/2)}{(q_n h/2)}.
\end{equation}
The fields for ferromagnetic Fe (having $\mu_0 M_s$=2.2 T) are shown in figure \ref{figS2}. We find
that the decay of the field strength with distance from the ferromagnet is determined by $\lambda$,
and that for reasonable parameters the local magnetostatic fields can easily be in the range of 1
kOe to 100 Oe up to a distance of 10 nm away from the surface of the ferromagnet, thus having a
significant impact on the spin accumulation and spin dynamics near the interface.

\section{Roughness characterization of devices}
\indent Since the magnitude of the local magnetostatic field near a ferromagnetic interface depends
on the amplitude and lateral period of the roughness, we performed characterization of the
roughness using atomic force microscopy (AFM) under ambient conditions for some of the devices
(Fig. \ref{figS3}). The top panel shows an AFM image of the surface of the Al$_2$O$_3$ tunnel
barrier on p-type Si, prior to deposition of the metal electrode. The root-mean-square (RMS)
roughness is about 0.2 nm. An example of a cross-sectional height profile (right) reveals that the
peak-to-peak roughness $h$ is about 0.5 nm, while the lateral variation has two different length
scales of about 20 nm and 60 nm, respectively. This roughness is then copied to the bottom surface
of the ferromagnetic metal that is grown on top of the tunnel barrier. The observed roughness can
certainly cause local magnetostatic fields in the range of 1 kOe to 100 Oe up to a distance of 10
nm away from the surface of the ferromagnet.\\
\indent Because magnetostatic fields are long range and the ferromagnet is a thin film (thickness
$\sim$10nm), the local magnetostatic fields that penetrate into the semiconductor are not only
determined by the roughness of the bottom interface of the ferromagnet with the Al$_2$O$_3$ tunnel
barrier, but also by the roughness of the top surface of the ferromagnetic layer. Unfortunately,
oxidation of the ferromagnet's surface prevents a good {\em ex-situ} measurement under ambient
conditions, and hence no data on this is available. For the sake of completeness we did perform AFM
analysis of the top of the complete metal electrode stack, consisting of 10 nm Ni and a 10 nm Au
cap layer (bottom panels of figure \ref{figS3}), although this may not be representative of the
roughness of the top surface of the FM. The roughness amplitude is significantly larger (RMS
roughness of 1.4 nm and a peak-to-peak amplitude of 3-4 nm) compared to the surface of the tunnel
barrier, while there is no small scale (20 nm) lateral roughness.

\end{appendix}
\section{Acknowledgements}
\indent The authors are grateful to A. Fert for continuing support,
discussions and feedback, to M. Tran for his contribution to the GaAs device development, and to M.
Yamamoto and T. Yorozu for the roughness characterization by AFM. This work was financially
supported by the Netherlands Foundation for Fundamental Research on Matter (FOM).


\clearpage

\begin{figure}[htb]
\includegraphics*[width=88mm]{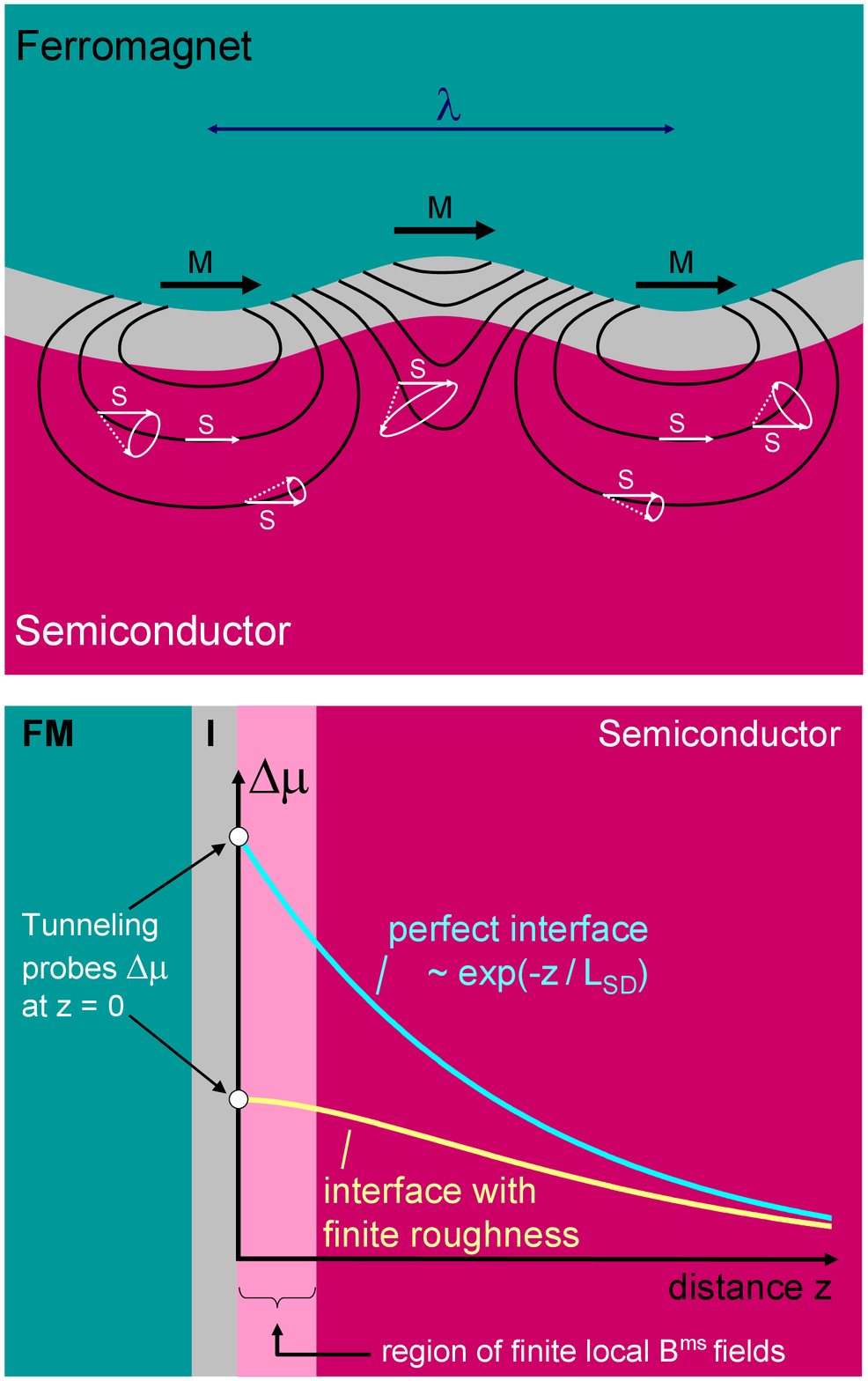}
\vspace*{0mm}\caption{{\bf Illustration of local interface magnetic fields and their effect on spin
precession in a semiconductor.} {\bf a,} The inhomogeneous magnetostatic field near a ferromagnetic
interface with finite roughness, sketched for a sinusoidal interface profile with period $\lambda$.
Field lines are in black, the magnetization of the ferromagnet (black arrows) points strictly along
the global interface plane everywhere. Spins are injected into the semiconductor with spin
initially aligned with the magnetization of the ferromagnet (solid white arrows). In the local
fields, the spins are precessing on different trajectories represented by dotted arrows and white
ellipses. Also the strength of the local field and hence the precession frequency is spatially
inhomogeneous. {\bf b,} Decay of the spin accumulation $\Delta \mu$ as a function of distance $z$
from the oxide/semiconductor interface for (i) a perfectly smooth interface (exponential decay with
spin-diffusion length $L_{SD}$), and (ii) an interface with finite roughness. For the latter, the
region in which the local magnetostatic fields B$^{ms}$ have an appreciable value is given in pink.
Note that tunneling probes the value of $\Delta \mu$ at $z=0$.} \label{fig1}
\end{figure}

\begin{figure}[htb]
\includegraphics*[width=180mm]{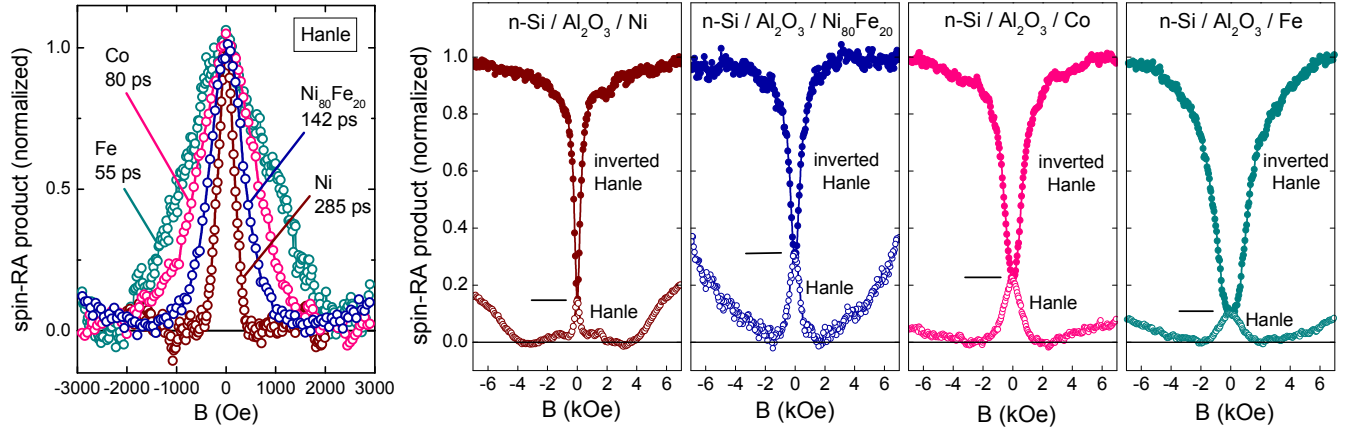}
\vspace*{5mm}\caption{{\bf Spin accumulation and precession in n-type silicon near a ferromagnetic
interface.} Room temperature data for n-Si/Al$_2$O$_3$/ferromagnet junctions with Ni,
Ni$_{80}$Fe$_{20}$, Co or Fe electrode. The vertical axis gives the spin-RA product, defined as
($\Delta$V/I)$\times$area. The magnetic field is applied perpendicular to the interface plane (open
symbols, Hanle), or parallel to the interface (solid symbols, inverted Hanle), with V$_{Si}-V_{FM}$
= +172 mV (electron injection). In the left panel, Hanle curves for different FM are normalized for
better comparison of the line width, denoted by an {\em effective} time $1/\omega$ representing the
width at half maximum of a fit to a Lorentzian (using g$=$2).} \label{fig2}
\end{figure}

\begin{figure}[htb]
\includegraphics*[width=180mm]{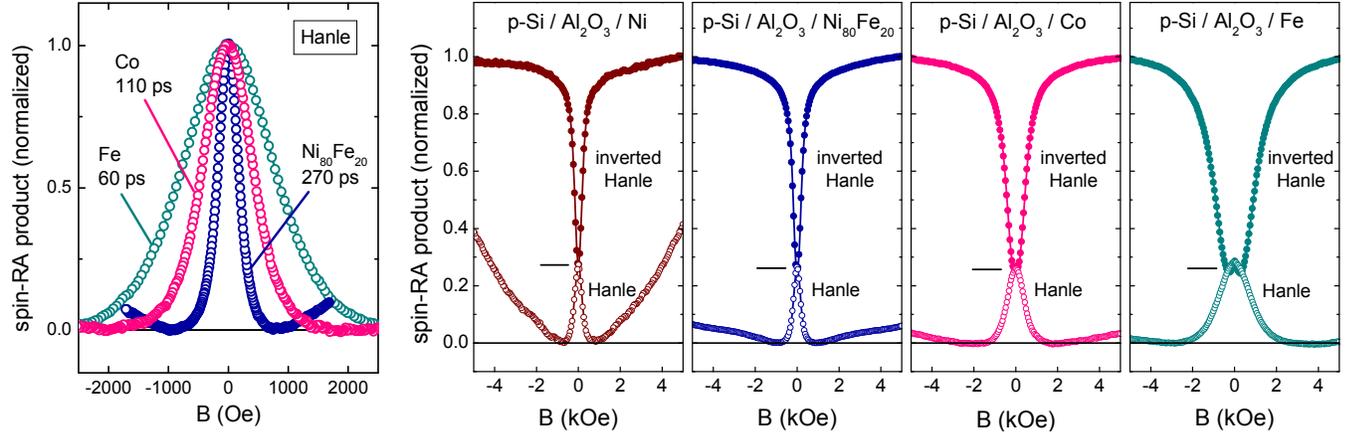}
\vspace*{5mm}\caption{{\bf Spin accumulation and precession in p-type silicon near a ferromagnetic
interface.} Room temperature data for p-Si/Al$_2$O$_3$/ferromagnet junctions with Ni,
Ni$_{80}$Fe$_{20}$, Co or Fe electrode. The magnetic field is applied perpendicular to the
interface plane (open symbols, Hanle), or parallel to the interface (solid symbols, inverted
Hanle), with V$_{Si}-V_{FM}$ = -172 mV (hole injection). In the left panel, Hanle curves for
different FM are normalized for better comparison of the line width, denoted by an {\em effective}
time $1/\omega$ representing the width at half maximum of a fit to a Lorentzian (using g$=$2).}
\label{fig3}
\end{figure}

\begin{figure}[htb]
\includegraphics*[width=88mm]{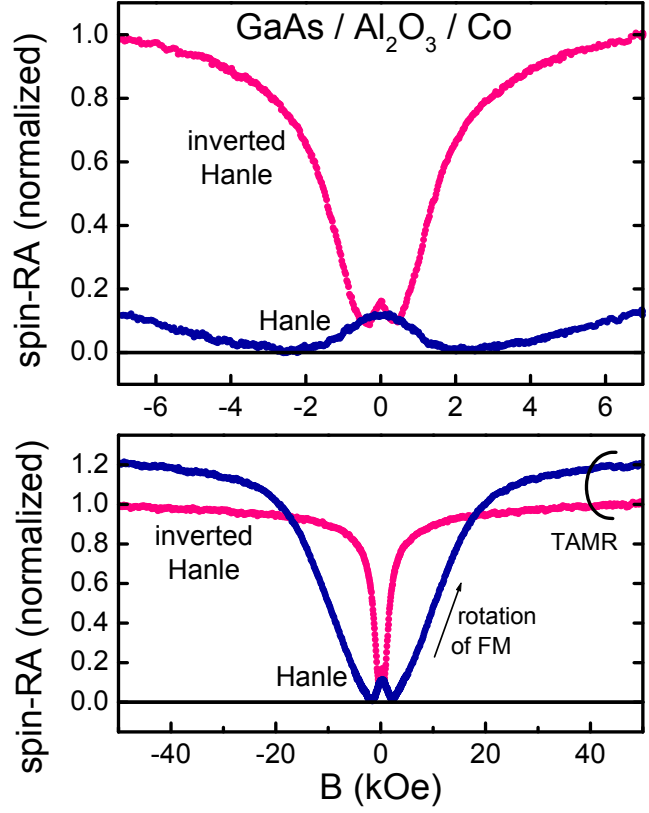}
\vspace*{0mm}\caption{{\bf Spin accumulation and precession in GaAs near a ferromagnetic
interface.} Experimental data for n-type GaAs/Al$_2$O$_3$/Co structures at 10 K, for magnetic field
applied perpendicular to the interface plane (blue, Hanle), or parallel to the interface (pink,
inverted Hanle). Data at V$_{GaAs}-V_{Co}$ = +422 mV (top panel) and +580 mV (bottom panel).}
\label{fig4}
\end{figure}

\begin{figure}[htb]
\includegraphics*[width=88mm]{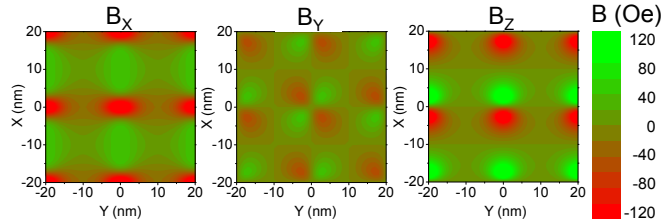}
\vspace*{0mm}\caption{{\bf Profiles of magnetostatic fields near a ferromagnetic interface with
finite roughness.} Calculated B$_x$, B$_y$ and B$_z$ components of the field versus position in the
x-y plane at 5 nm distance from the interface, represented by a 2D square array of magnetic dipoles
pointing along the x-axis (period $\lambda =$ 20 nm, central dipole at x = y = 0). The dipole
strength $\mu$ is such that $\mu_0\mu/4\pi =$ 2 Tnm$^{3}$.} \label{fig5}
\end{figure}

\begin{figure}[htb]
\includegraphics*[width=180mm]{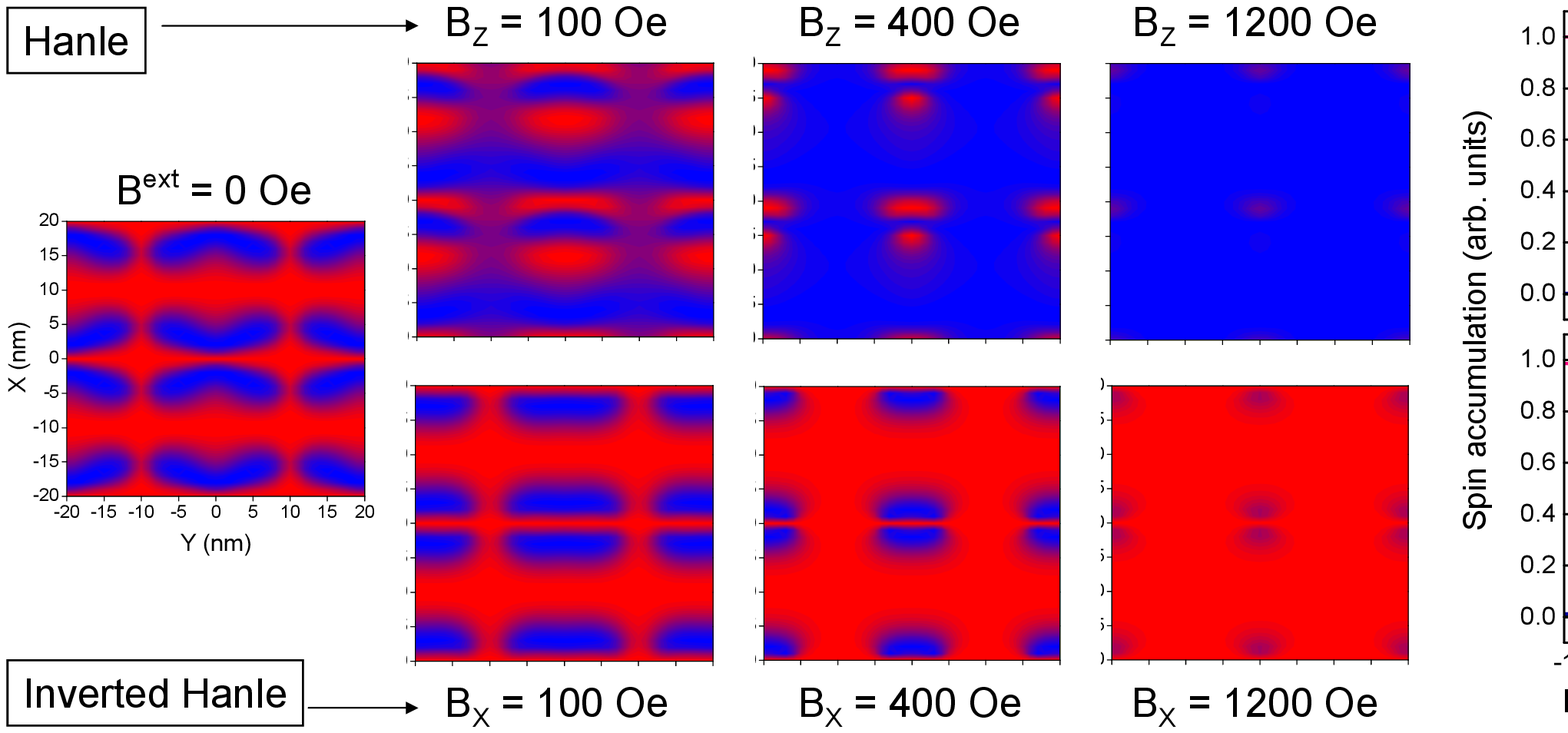}
\vspace*{0mm}\caption{{\bf Calculated spin accumulation near an interface with local magnetostatic
fields.} Maps of the S$_x$ component of the spin density in the semiconductor versus position in
the x-y plane parallel to the interface, for different values of external magnetic field applied
along the z-axis (Hanle configuration, top row), or along the x-axis parallel to the magnetization
of the ferromagnetic injector (inverted Hanle configuration, bottom row). Red color corresponds to
the maximum spin density (without any precession), blue to zero spin accumulation. The
magnetostatic fields were taken at 5 nm distance from a dipole array with $\lambda=$ 20 nm and
$\mu_0\mu/4\pi =$ 10 Tnm$^{3}$. The spin lifetime was set to 1 ns. Right panels show the resulting
Hanle (blue) and inverted Hanle (pink) curves for external applied magnetic field along z or x
axis, respectively, calculated by averaging the inhomogeneous spin density over the x-y plane. The
dipole strength is such that $\mu_0\mu/4\pi = 2$ (top panel) or 10 Tnm$^{3}$ (bottom panel). Also
shown in green are pure Lorentzian line shapes for the same 1 ns spin lifetime, with the peak
amplitude scaled for easy comparison.} \label{fig6}
\end{figure}

\begin{figure}[htb]
\includegraphics*[width=88mm]{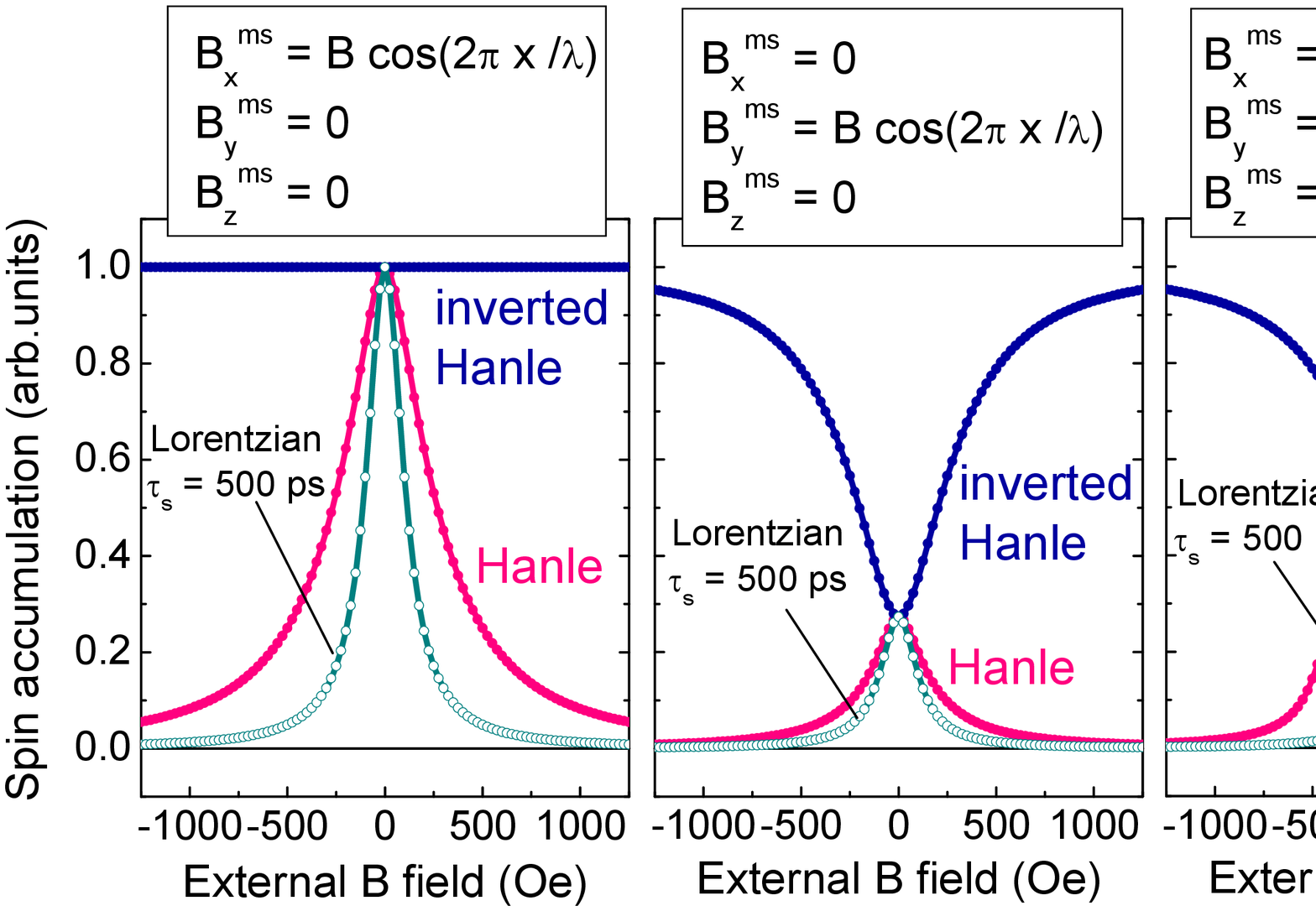}
\vspace*{0mm}\caption{Hanle (pink) and inverted Hanle (blue) curves for external applied magnetic
field B$^{ext}$ along z or x axis, respectively, calculated using eqn. (11). Included are
magnetostatic fields B$^{ms}$ pointing purely along either the x, y or z-axis, as indicated, and
with a strength that has a simple sinusoidal spatial variation. The spin lifetime $\tau_s$ was set
to 500 ps. Also shown in green are for the same 500 ps spin lifetime the pure Lorentzian curves,
with the amplitude adjusted for easy comparison.} \label{figS1}
\end{figure}

\begin{figure}[htb]
\includegraphics*[width=88mm]{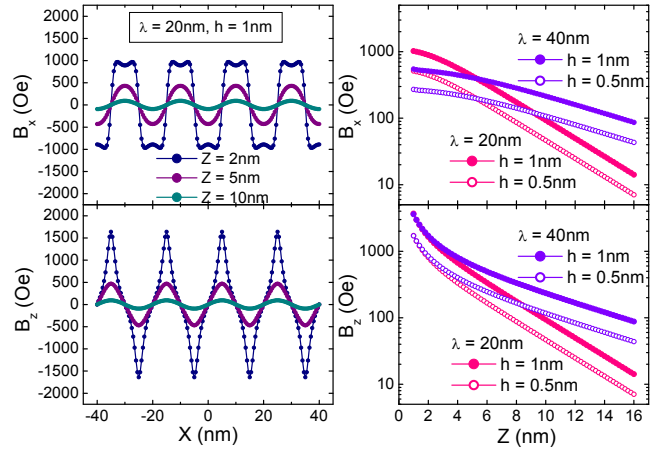}
\vspace*{0mm}\caption{Left panels: Calculated strength of the B$_x$ (top) and B$_z$ (bottom)
component of the local magnetostatic field as a function of lateral $x$-position at different
distance $z$ from a ferromagnetic Fe surface with 1-dimensional roughness, for $\lambda$=20nm and
roughness amplitude $h$=1nm. Right panels: the same, but now as a function of distance $z$ at fixed
$x$-position, for $\lambda$=20nm or 40nm and roughness amplitude $h$=1nm or 0.5nm.} \label{figS2}
\end{figure}

\begin{figure}[htb]
\includegraphics*[width=88mm]{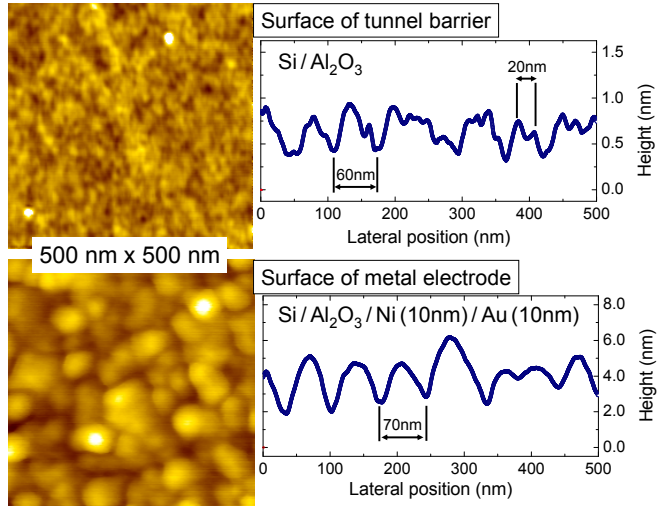}
\vspace*{0mm}\caption{Top panels: Atomic force image (500 nm $\times$ 500 nm) of the surface of the
Al$_2$O$_3$ tunnel barrier on p-type Si, prior to metal electrode deposition, and a representative
cross-sectional height profile. Bottom panels: the same, but now after deposition of the
ferromagnetic metal electrode (Ni, 10 nm) and the Au cap layer (10 nm).} \label{figS3}
\end{figure}

\end{document}